\documentclass[12pt]{iopart}

\usepackage{graphicx,color}
\usepackage{epsfig}
\usepackage{wrapfig}

\begin{document}

\title[Wibson W. G. Silva and José Holanda...................................................................]{One Analytical Approach of Rashba-Edelstein Magnetoresistance in 2D Materials}

\author{Wibson W. G. Silva and José Holanda$^{*}$}

\address{Programa de Pós-Graduação em Engenharia Física, Universidade Federal Rural de Pernambuco, 54518-430, Cabo de Santo Agostinho, Pernambuco, Brazil}

\ead{$^{*}$joseholanda.silvajunior@ufrpe.br}

\vspace{10pt}

\vspace{1cm}
\begin{abstract}
We study analytically the Rashba-Edelstein magnetoresistance (REMR) in a structure made from an insulator ferromagnet, such as yttrium iron garnet (YIG), and a 2D material (2DM) with direct and inverse Rashba-Edelstein effects, such as SLG and MoS$_2$. Our results represent an efficient way of analyzing the Rashba-Edelstein effects.
\end{abstract}
\vspace{1cm}
\begin{indented}
	\item[]To ArXiv 
\end{indented}

\vspace{1cm}
\section{Introduction}
{\fontsize{15}{\baselineskip} \textbf{2D}} spintronics has gained an important meaning in data storage technologies, in many those cases, the 2D materials are non-magnetic, however, magnetism can be induced at the interface of those materials. Among some methods, two of them are broadly applied to induce magnetism on 2D material. The first method is to introduce vacancies or adding atoms producing spin polarization \cite{Soumyanarayanan, Hellman, Han}. The other one is to induce magnetism of the adjacent magnetic materials via the magnetic proximity effect \cite{Rojas, Leutenantsmeyer, YWangY, McCreary}. Recently was discovered that 2D magnetic van der Waals crystals have intrinsic magnetic ground states at the atomic scale, providing new opportunities in the field of 2D spintronics \cite{Chang, Wei}. Furthermore, it was discovered that several materials as single-layer graphene (SLG) and molybdenum disulfide (MoS$_2$) \cite{Wang, Tang, Gong} can also be used for spin-charge current conversion \cite{Zhong, Mak, Palacios, Amani, Aparecido, Zhang}. Due to their layered structures, MoS$_2$ and SLG can be easily prepared with one or several atomic layers to explore the transport properties. SLG and semiconducting MoS$_2$ have 2D electronic states that are expected to exhibit remarkable pseudospin and spin-momentum locking, respectively \cite{Wang, Shao, Almeida, Pesin, Alves}. These are essential ingredients for the charge-to-spin current conversion by the direct Rashba-Edelstein effect (REE) or for spin-to-charge current conversion by the inverse Rashba Edelstein effect (IREE). Another fundamental ingredient is the broken inversion symmetry at material surfaces and interfaces \cite{Soumyanarayanan, Hellman, Han, Vidyasagar, Sanchez, Loreto, Mori, Holanda}.

Although the change of electrical resistance of ferromagnets has been studied for a long time, providing a fundamental understanding of spin-dependent transport in different structures \cite{Sanchez, Loreto, Mori}, the transport properties of 2D materials still present themselves as a challenge. One of the most important effects in spin-dependent transport is the spin Hall magnetoresistance (SMR) \cite{Holanda, Chen, Mckenzie}. In 3D materials, the SMR is explained by the spin-current reflection and reciprocal spin-charge conversion caused by the simultaneous action of the spin Hall effect (SHE) \cite{JHolanda,Nakayama,Althammer} and inverse spin Hall effect (ISHE) \cite{Takahashi}. The challenge is to explore the magnetoresistance induced in 2D materials \cite{Conductiond, Nomura}. In this paper, we present a study based on direct and inverse Rashba-Edelstein effects that describes the magnetoresistance in 2D materials, which is called of Rashba-Edelstein magnetoresistance (REMR).

\section{2D materiais in contact with a magnetic insulator}

The REMR is induced by the simultaneous action of direct and inverse Rashba-Edelstein effects and therefore a nonequilibrium proximity phenomenon. The magnetoresistance study was carried out with arrangement as illustrated in Fig. \ref{puga} below.
\begin{figure}[h]
\vspace{0.1mm} \hspace{0.1mm}
\begin{center}
\includegraphics[scale=0.3]{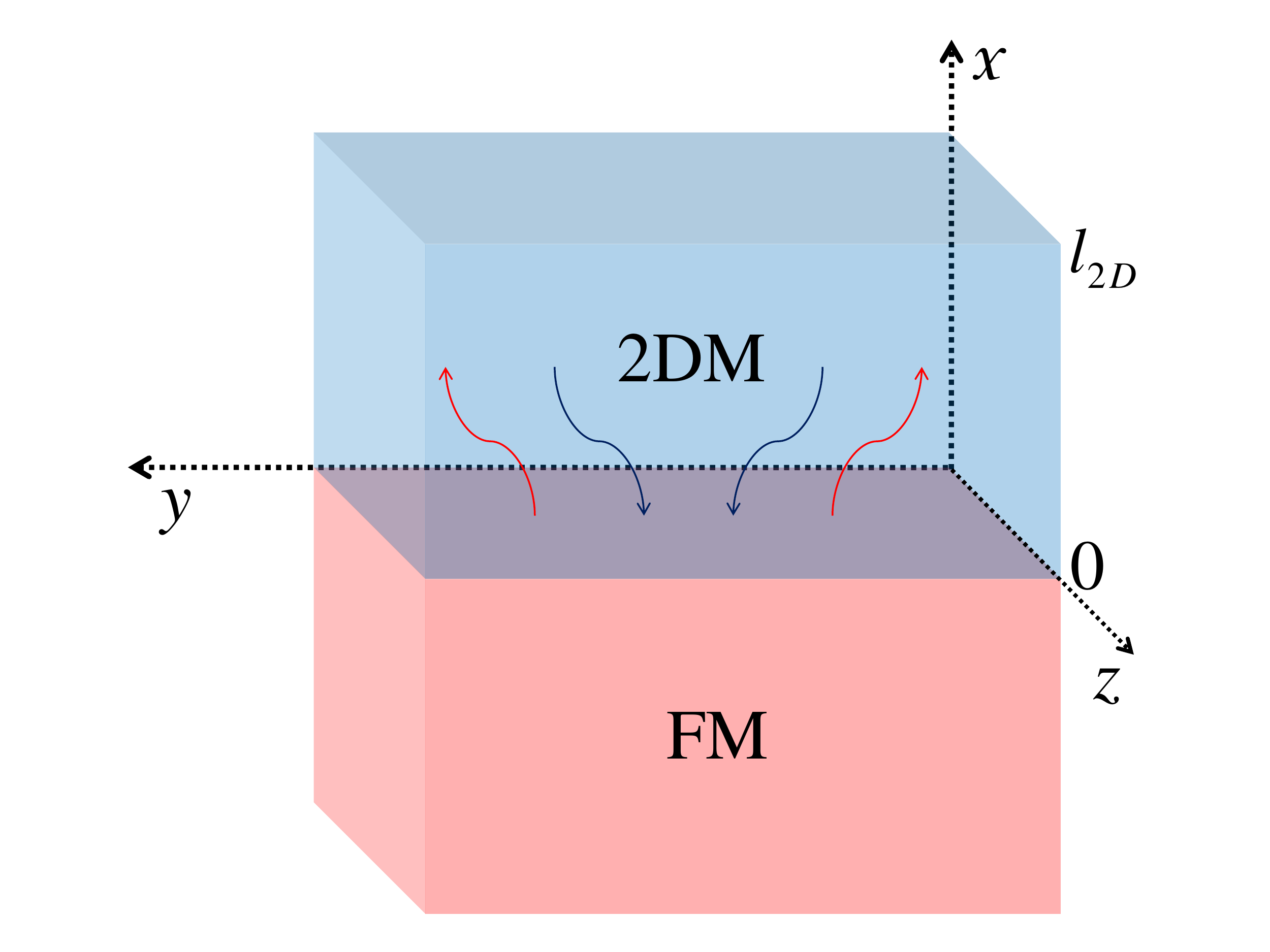}
\caption{\label{arttype}(online color) Illustration of the sample structure used to study the Rashba-Edelstein magnetoresistance (REMR).}

\label{puga}
\end{center}
\end{figure}
The effects of the spin current in 2D materials are very important for phenomena of transport. Considering the Ohm’s law for 2D materials with direct and inverse Rashba-Edelstein effects and therefore a nonequilibrium proximity phenomenon can be understood by the relation between thermodynamic driving force and currents that reflects Onsager’s reciprocity by the symmetry of the response matrix: 

\begin{equation} 
\left(\begin{array}{cc}
	\vec{J}_{C}\\
	\vec{J}_{Sx}\\
	\vec{J}_{Sy}\\
	\vec{J}_{Sz} 
\end{array}\right) = \frac{1}{R_{2D}}
\left(\begin{array}{cccc}
		1 & \hat{x} \times & \hat{y} \times & \hat{z} \times\\
		\frac{1}{\lambda_{REE}} \hat{x} \times  & \frac{1}{\lambda_{REE}} & 0 & 0\\
		\frac{1}{\lambda_{REE}} \hat{y} \times & 0 & \frac{1}{\lambda_{REE}} & 0 \\
		\frac{1}{\lambda_{REE}} \hat{z} \times & 0 & 0 & \frac{1}{\lambda_{REE}}
\end{array}\right)
\left(\begin{array}{cc}
	-\nabla \mu_{C}/e\\
	-\nabla \mu_{Sx}/2e\\
	-\nabla \mu_{Sy}/2e\\
	-\nabla \mu_{Sz}/2e 
\end{array}\right),
\label{1}
\end{equation}
where \textit{e} = $\mid e \mid$ is the electron charge, R$_{2D}$ is the resistance of 2D material, $\mu_{C}$ is the charge chemical potential, $\vec{\mu}_{S}$ is the spin accumulation, $\vec{J}_{C}$ is the charge current density and $\vec{J}_{S}$ is the spin current density. The direct Rashba-Edelstein is represented by the lower diagonal elements that generate the spin currents in the presence of an applied current density, which generates an electric field, in the following chosen to be in the $\hat{x}$ direction $\vec{E} = E_{x}\hat{x} = -\hat{x}(\partial_{x}\mu_{C}/e)$. On the other hand, the inverse Rashba-Edelstein effect is governed by element above the diagonal that connect the gradients of the spin accumulations to the charge current density. The spin accumulation $\vec{\mu}_{S}$ is obtained from the spin-diffusion equation in the 2D materials      

\begin{equation} 
\nabla^{2}\vec{\mu_{S}} = \frac{\vec{\mu}_{S}}{\lambda_{SD}^{2}},
	\label{2}
\end{equation}
where $\lambda_{SD}$ is the spin-diffusion length. Spin accumulation is always due to spin diffusion, which even for a 2D material such as graphene has spin diffusion in the z-direction. For 2D materials with thickness $l_{2D}$ in the $\hat{x}$ direction the solution of equation (2) is

\begin{equation} 
	\vec{\mu}_{S}(z) = \vec{p}\e^{-z/\lambda_{2D}} + \vec{q	}\e^{z/\lambda_{2D}},
	\label{3}
\end{equation}
where the constant column vectors $\vec{p}$ and $\vec{q}$ are determined by the boundary conditions at the interfaces. According to Eq. (2), the spin current in 2D materials consists of spin diffusion process. For a system homogeneous in the \textit{x-y} plane, the spin current density flowing in the $\hat{z}$ direction is 

\begin{equation} 
\vec{J}_{S}^{z}(z) = - \left(\frac{1}{2eR_{2D}\lambda_{REE}}\right)\partial_{z}\vec{\mu}_{Sz} - J_{SO}^{REE}\hat{y},
	\label{4}
\end{equation}
where $J_{SO}^{REE} = E_{x}/R_{2D}\lambda_{REE}$ is the bare Rashba-Edelstein current, i. e., the spin current generated directly by the REE and $\lambda_{REE}$ is the REE length. At the interfaces $z = l_{2D}$ and \textit{z} = 0 the boundary conditions demand that $\vec{J}_{S}^{z}(z)$ is continuous. The spin current at $z = l_{2D}$ interface vanishes, $\vec{J}_{S}^{z}(z = l_{2D}) = \vec{J}_{S}^{2D} = 0$. On the other hand, in general at the magnetic interface the spin current density $\vec{J}_{S}^{FM}$ is governed by the spin accumulation and spin-mixing conductance \cite{Adachi}, such that:       

\begin{equation} 
	\vec{J}_{S}^{FM}(\hat{m}) = g_{r}\hat{m}\times\left(\hat{m} \times \frac{\vec{\mu}_{S}}{e}\right) + g_{i}\left(\hat{m} \times \frac{\vec{\mu}_{S}}{e}\right),
	\label{5}
\end{equation}
where $\hat{m} = (m_{x}, m_{y}, m_{z})^T$ represents a unit vector along the magnetization and $g_{\uparrow\downarrow} = g_{r} + ig_{i}$ the complex spin-mixing interface conductance per unit length and resistance. It is agreed that $g_r$ characterizes the efficiency of the interfacial spin transport and the imaginary part $g_i$ can be interpreted as an effective exchange field acting on the spin accumulation. According with Eq. (5) a positive current corresponds to up spins moving from FM towards 2D. In particular for FM insulator, this spin current density is proportional to the spin transfer torque acting on the ferromagnet

\begin{equation} 
	\vec{\tau}_{STT} = -\frac{\hbar}{2e} \hat{m}\times\left(\hat{m} \times \vec{J}_{S}^{FM} \right) = \frac{\hbar}{2e} \vec{J}_{S}^{FM}(\hat{m}).
	\label{6}
\end{equation}
Using the boundary conditions discussed before, it is possible to determine the coefficients $\vec{p}$ and $\vec{q}$, which leads to the spin accumulation for structures 2DM/FM  

\begin{equation*} 
\vec{\mu}_{S}(z) = - \mu_{SO} \left[\frac{sinh\left(\frac{2z - l_{2D}}{2\lambda_{2D}}\right)}{sinh\left(\frac{l_{2D}}{2\lambda_{2D}}\right)}\right]\hat{y} + 
\end{equation*}
\begin{equation}
2e\lambda_{2D}\lambda_{REE}R_{2D}\left[\frac{cosh\left(\frac{z - l_{2D}}{2\lambda_{2D}}\right)}{sinh\left(\frac{l_{2D}}{2\lambda_{2D}}\right)}\right]\vec{J}_{S}^{FM}(\hat{m})
	\label{7}
\end{equation}
where $\mu_{SO} \equiv |\vec{\mu}_{S}(0)| = 2e\lambda_{2D}\lambda_{REE}R_{2D}J_{SO}^{REE}tanh(l_{2D}/2\lambda_{2D})$ is the spin accumulation at the interface in the absence of spin transfer, i. e., when $g_{\uparrow\downarrow} = 0$. Furthermore, according with Eq. (5), the spin accumulation at \textit{z} = 0 becomes 

\begin{equation*} 
	\vec{\mu}_{S}(0) = \mu_{SO}\hat{y}  + 2\lambda_{2D}\lambda_{REE}R_{2D}coth\left(\frac{l_{2D}}{\lambda_{2D}}\right) \times
\end{equation*}
\begin{equation}
	\left[g_{r}\left\{\hat{m}\left(\hat{m}\cdot\vec{\mu}_{S}(0)\right)-\vec{\mu}_{S}(0)\right\}+g_{i}\hat{m} \times \vec{\mu}_{S}(0)\right] 
	\label{8}
\end{equation}
where $\hat{m}\cdot\vec{\mu}_{S}(0)$ and $\hat{m} \times \vec{\mu}_{S}(0)$ here are

\begin{equation}
	\hat{m}\cdot\vec{\mu}_{S}(0) = m_{y}\mu_{SO}, 
	\label{9}
\end{equation}

\begin{equation*}
	\hat{m} \times \vec{\mu}_{S}(0) = \mu_{SO}\left[\frac{\left(\frac{\hat{m} \times \hat{y}}{R_{2D}\lambda_{REE}}\right)+\left(2m_{y}\lambda_{2D}g_{i}coth\left(\frac{l_{2D}}{\lambda_{2D}}\right)\right)\hat{m}}{\frac{1}{R_{2D}\lambda_{REE}}+2\lambda_{2D}g_{r}coth\left(\frac{l_{2D}}{\lambda_{2D}}\right)}\right] 
	\label{10}
\end{equation*}
\begin{equation}
	-\left[\frac{2\lambda_{2D}g_{i}coth\left(\frac{l_{2D}}{\lambda_{2D}}\right)}{\frac{1}{R_{2D}\lambda_{REE}}+2\lambda_{2D}g_{r}coth\left(\frac{l_{2D}}{\lambda_{2D}}\right)}\right]\vec{\mu}_{S}(0), 
	\label{11}
\end{equation}
and
\begin{equation}
	\vec{\mu}_{S}(0) = \mu_{SO}\left[\frac{(A(1+A)+B^2)\hat{m}+B(\hat{m} \times \hat{y})+(1+A)\hat{y}}{A^2 + B^2}\right],
	\label{12}
\end{equation}
where
\begin{equation}
	A = 2\lambda_{2D}\lambda_{REE}R_{2D}g_{r}coth\left(\frac{l_{2D}}{\lambda_{2D}}\right)
	\label{13}
\end{equation}
and
\begin{equation}
	B = 2\lambda_{2D}\lambda_{REE}R_{2D}g_{i}coth\left(\frac{l_{2D}}{\lambda_{2D}}\right).
	\label{14}
\end{equation}
The spin current through the FM/2DM interfaces reads
\begin{equation*}
	\vec{J}_{S}^{FM} = \left(\frac{\mu_{SO}}{eR_{2D}\lambda_{REE}}\right)\left[Im\left\{\frac{g_{\uparrow\downarrow}}{\frac{1}{R_{2D}\lambda_{REE}}+2\lambda_{2D}g_{\uparrow\downarrow}coth\left(\frac{l_{2D}}{\lambda_{2D}}\right)}\right\}\right]\left(\hat{m} \times \hat{y}\right)
\end{equation*}
\begin{equation}
	+\left(\frac{\mu_{SO}}{eR_{2D}\lambda_{REE}}\right)\left[Re\left\{\frac{g_{\uparrow\downarrow}}{\frac{1}{R_{2D}\lambda_{REE}}+2\lambda_{2D}g_{\uparrow\downarrow}coth\left(\frac{l_{2D}}{\lambda_{2D}}\right)}\right\}\right]\hat{m} \times \left(\hat{m} \times \hat{y}\right).
	\label{14}
\end{equation}
In this way, the spin accumulation is,
\begin{equation*}
	\frac{\vec{\mu}_{S}(z)}{\mu_{SO}} = 
	Im\left[\left\{\frac{2\lambda_{2D}g_{\uparrow\downarrow}}{\frac{1}{R_{2D}\lambda_{REE}}+2\lambda_{2D}g_{\uparrow\downarrow}coth\left(\frac{l_{2D}}{\lambda_{2D}}\right)}\right\}\left\{\frac{cosh\left(\frac{z-l_{2D}}{\lambda_{2D}}\right)}{sinh\left(\frac{l_{2D}}{\lambda_{2D}}\right)}\right\}\right](\hat{m} \times 
\end{equation*}
\begin{equation*}
	\hat{y})+Re\left[\left\{\frac{2\lambda_{2D}g_{\uparrow\downarrow}}{\frac{1}{R_{2D}\lambda_{REE}}+2\lambda_{2D}g_{\uparrow\downarrow}coth\left(\frac{l_{2D}}{\lambda_{2D}}\right)}\right\}\left\{\frac{cosh\left(\frac{z-l_{2D}}{\lambda_{2D}}\right)}{sinh\left(\frac{l_{2D}}{\lambda_{2D}}\right)}\right\}\right]\hat{m} \times (\hat{m} \times 
\end{equation*}
\begin{equation}
	\hat{y})-\left[\frac{sinh\left(\frac{2z-l_{2D}}{2\lambda_{2D}}\right)}{sinh\left(\frac{2z-l_{2D}}{2\lambda_{2D}}\right)}\right]\hat{y},
	\label{15}
\end{equation}
then leads to the distributed spin current in 2DM
\begin{equation*}
	\frac{\vec{J}_{S}^{z}(z)}{J_{SO}^{REE}} = 
	-Im\left[\left\{\frac{2\lambda_{2D}g_{\uparrow\downarrow}}{\frac{1}{R_{2D}\lambda_{REE}}+2\lambda_{2D}g_{\uparrow\downarrow}coth\left(\frac{l_{2D}}{\lambda_{2D}}\right)}\right\}\left\{\frac{sinh\left(\frac{z-l_{2D}}{\lambda_{2D}}\right)}{sinh\left(\frac{l_{2D}}{\lambda_{2D}}\right)}\right\}\right](\hat{m} \times 
\end{equation*}
\begin{equation*}
 \hat{y})-Re\left[\left\{\frac{2\lambda_{2D}g_{\uparrow\downarrow}}{\frac{1}{R_{2D}\lambda_{REE}}+2\lambda_{2D}g_{\uparrow\downarrow}coth\left(\frac{l_{2D}}{\lambda_{2D}}\right)}\right\}\left\{\frac{sinh\left(\frac{z-l_{2D}}{\lambda_{2D}}\right)}{sinh\left(\frac{l_{2D}}{\lambda_{2D}}\right)}\right\}\right]\hat{m} \times (\hat{m} \times 
\end{equation*}
\begin{equation}
	\hat{y})+\left[\frac{cosh\left(\frac{2z-l_{2D}}{2\lambda_{2D}}\right)-cosh\left(\frac{l_{2D}}{2\lambda_{2D}}\right)}{cosh\left(\frac{l_{2D}}{2\lambda_{2D}}\right)}\right]\hat{y}.
	\label{16}
\end{equation}
The inverse Rashba-Edelstein effect drives a charge current in the \textit{x-y} plane by the diffusion spin current component flowing along the \textit{z} direction. The total longitudinal (along $\hat{x}$) component is  
\begin{equation*}
	\frac{J_{C, long}(z)}{J_{CO}} = 1+4\left(\frac{\lambda_{REE}}{l_{2D}}\right)^2
	\left[\frac{cosh\left(\frac{z-l_{2D}}{\lambda_{2D}}\right)}{cosh\left(\frac{l_{2D}}{\lambda_{2D}}\right)}+\left(1-m_{y}^{2}\right)\right] \times	
\end{equation*}
\begin{equation}
	Re\left[\left\{\frac{2\lambda_{2D}g_{\uparrow\downarrow }tanh\left(\frac{l_{2D}}{2\lambda_{2D}}\right)}{\frac{1}{R_{2D}\lambda_{REE}}+2\lambda_{2D}g_{\uparrow\downarrow}coth\left(\frac{l_{2D}}{\lambda_{2D}}\right)}\right\}\left\{\frac{sinh\left(\frac{z-l_{2D}}{\lambda_{2D}}\right)}{sinh\left(\frac{l_{2D}}{\lambda_{2D}}\right)}\right\}\right]
	\label{17}
\end{equation}
aand transverse or Rashba-Edelstein component is
\begin{equation*}
	\frac{J_{C, trans}(z)}{J_{CO}} = 4\left(\frac{\lambda_{REE}}{l_{2D}}\right)^2
	\left[m_{x}m_{y}Re-m_{y}Im \right] \times	
\end{equation*}
\begin{equation}
	Re\left[\left\{\frac{2\lambda_{2D}g_{\uparrow\downarrow }tanh\left(\frac{l_{2D}}{2\lambda_{2D}}\right)}{\frac{1}{R_{2D}\lambda_{REE}}+2\lambda_{2D}g_{\uparrow\downarrow}coth\left(\frac{l_{2D}}{\lambda_{2D}}\right)}\right\}\left\{\frac{sinh\left(\frac{z-l_{2D}}{\lambda_{2D}}\right)}{sinh\left(\frac{l_{2D}}{\lambda_{2D}}\right)}\right\}\right]
	\label{18}
\end{equation}
where $J_{CO} = E_{x}/(R_{2D}\lambda_{REE})$ it is the charge current driven by the external electric current. Expanding the longitudinal resistance governed by the current in the \textit{x}-direction of the applied field to leading order in $\lambda_{REE}^{2}$ and averaging the electric currents over the 2DM thickness, it is finded
\begin{equation}
\left(R_{2D}\right)_{long} = \left(\frac{\lambda_{REE}E_{x}}{\overline{J_{C, long}}}\right) \approx R_{2D}+\Delta R_{2D}^{(0)}+(1-m_{y}^{2})\Delta R_{2D}^{(1)},
\label{19}
\end{equation}
and
\begin{equation}
	\left(R_{2D}\right)_{trans} \approx -\left(\frac{\overline{J_{C, trans}}}{E_x}\right) \left(\frac{1}{R_{2D}\lambda_{REE}}\right)^2 = m_{x}m_{y}\Delta R_{2D}^{(1)} + m_{z}\Delta R_{2D}^{(2)},
	\label{20}
\end{equation}
where
\begin{equation}
	\frac{\Delta R_{2D}^{(0)}}{R_{2D}} = - \left(\frac{2\lambda_{REE}}{l_{2D}}\right)^2\left(\frac{2\lambda_{2D}}{l_{2D}}\right)tanh\left(\frac{l_{2D}}{2\lambda_{2D}}\right),
	\label{21}
\end{equation}
\begin{equation}
	\frac{\Delta R_{2D}^{(1)}}{R_{2D}} = \left(\frac{2\lambda_{REE}}{l_{2D}}\right)^2\left(\frac{\lambda_{2D}}{l_{2D}}\right)Re\left[\frac{2\lambda_{2D}g_{\uparrow\downarrow }tanh^2\left(\frac{l_{2D}}{2\lambda_{2D}}\right)}{\frac{1}{R_{2D}\lambda_{REE}}+2\lambda_{2D}g_{\uparrow\downarrow}coth\left(\frac{l_{2D}}{\lambda_{2D}}\right)}\right],
	\label{22}
\end{equation}
\begin{equation}
	\frac{\Delta R_{2D}^{(2)}}{R_{2D}} = \left(\frac{2\lambda_{REE}}{l_{2D}}\right)^2\left(\frac{\lambda_{2D}}{l_{2D}}\right)Im\left[\frac{2\lambda_{2D}g_{\uparrow\downarrow }tanh^2\left(\frac{l_{2D}}{2\lambda_{2D}}\right)}{\frac{1}{R_{2D}\lambda_{REE}}+2\lambda_{2D}g_{\uparrow\downarrow}coth\left(\frac{l_{2D}}{\lambda_{2D}}\right)}\right]
	\label{23}
\end{equation}
and $\Delta R_{2D}^{(0)} < 0$, this suggests that the resistance is reduced by the Rashba interaction.

\section{Discussion and applications}

\subsection{Different 2D magnetoresistances}
\begin{figure}[h]
	\vspace{0.5mm} \hspace{0.5mm}
	\begin{center}
		\includegraphics[scale=0.3]{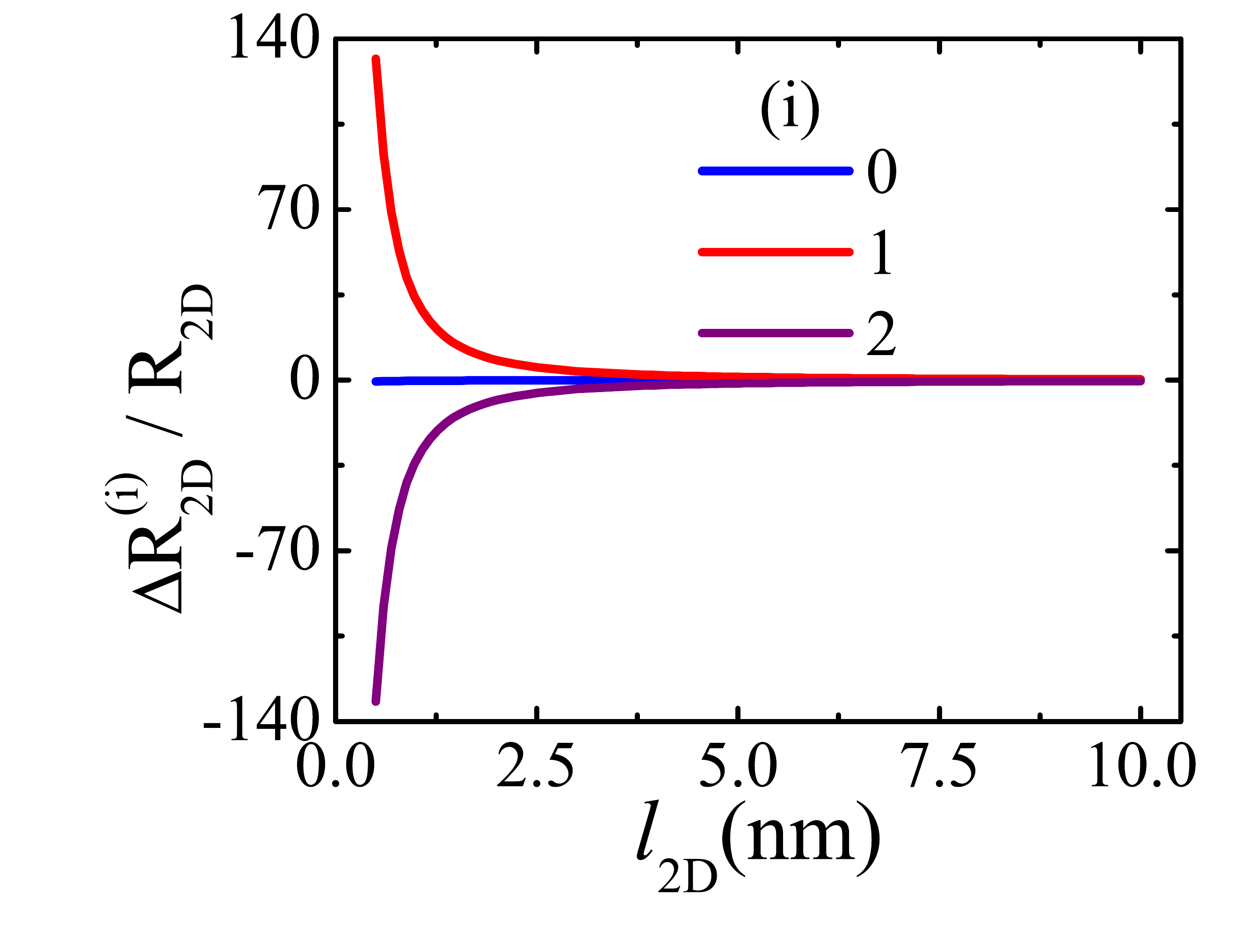}
		\caption{\label{arttype}(online color) 2D magnetoresistances $\Delta R_{2D}^{i}/R_{2D}$ as a function of thickness $l_{2D}$ with \textit{i} = 0, 1, 2, $g_{r} = 2.4 \times 10^7 $  m$^{-1} \Omega^{-1}$, $g_{i} = 2.4 \times 10^7$ m$^{-1} \Omega^{-1}$, $R_{2D} = 0.2 \times 10^6 $ $\Omega$, $\lambda_{2D} = 235 \times 10^{-9} $ m and $ \lambda_{REE} = 0.13 \times 10^{-9} $ m for MoS$_2$ \cite{Aparecido,Edelstein}.}
		\label{gato}
	\end{center}
\end{figure}

For small thickness (2D surface) $l_{2D} \ll \lambda_{2D}$ the equations (21), (22) and (23) are written as

\begin{equation}
	\frac{\Delta R_{2D}^{(0)}}{R_{2D}} = - \left(\frac{2\lambda_{REE}}{l_{2D}}\right),
	\label{24}
\end{equation}
\begin{equation}
	\frac{\Delta R_{2D}^{(1)}}{R_{2D}} = 2\left(\frac{\lambda_{REE}}{l_{2D}}\right)^2 \left[\frac{g_{r}R_{2D}\lambda_{REE}l_{2D}}{l_{2D}+2g_{r}R_{2D}\lambda_{REE}\lambda_{2D}^{2}}\right]
	\label{25}
\end{equation}
and
\begin{equation}
	\frac{\Delta R_{2D}^{(2)}}{R_{2D}} = - 2\left(\frac{\lambda_{REE}}{l_{2D}}\right)^2 \left[\frac{g_{i}R_{2D}\lambda_{REE}l_{2D}}{l_{2D}+2g_{i}R_{2D}\lambda_{REE}\lambda_{2D}^{2}}\right].
	\label{26}
\end{equation}
In Fig. \ref{gato}, it is shown the different 2D magnetoresistances $\Delta R_{2D}^{i}/R_{2D}$ as a function of thickness $l_{2D}$ with \textit{i} = 0, 1, 2, $g_{r} =  2.4 \times 10^7$ m$^{-1} \Omega^{-1}$, $g_{i} = 2.4 \times 10^7$ m$^{-1} \Omega^{-1}$, $R_{2D} = 0.215 \times 10^6 $ $\Omega$, $\lambda_{2D} = 235 \times 10^{-9} $ m and $ \lambda_{REE} = 0.13 \times 10^{-9} $ m for MoS$_2$ \cite{Aparecido,Edelstein}. The real ($g_{r}$) and imaginary ($g_{i}$) parts of the spin-mixing interface conductance were obtained considering the complex spin-mixing interface conductance module equal to the effective spin-mixing conductance obtained by spin pumping measurements \cite{Aparecido}. After we consider the dimensions of the structure, we get that the real ($g_{r}$) and imaginary ($g_{i}$) parts of the spin-mixing interface conductance are the same for the YIG/MoS$_2$ structure, which is an result extremely reasonable, considering the YIG /MoS$_2$ interface has both intrinsic spin-orbit coupling and proximity effect. The different behaviors described by 2D magnetoresistances in Fig. \ref{gato} reveal that are effects that can be measured differently and separately.  

\subsection{REE length}

\begin{figure}[h]
	\vspace{0.5mm} \hspace{0.5mm}
	\begin{center}
		\includegraphics[scale=0.3]{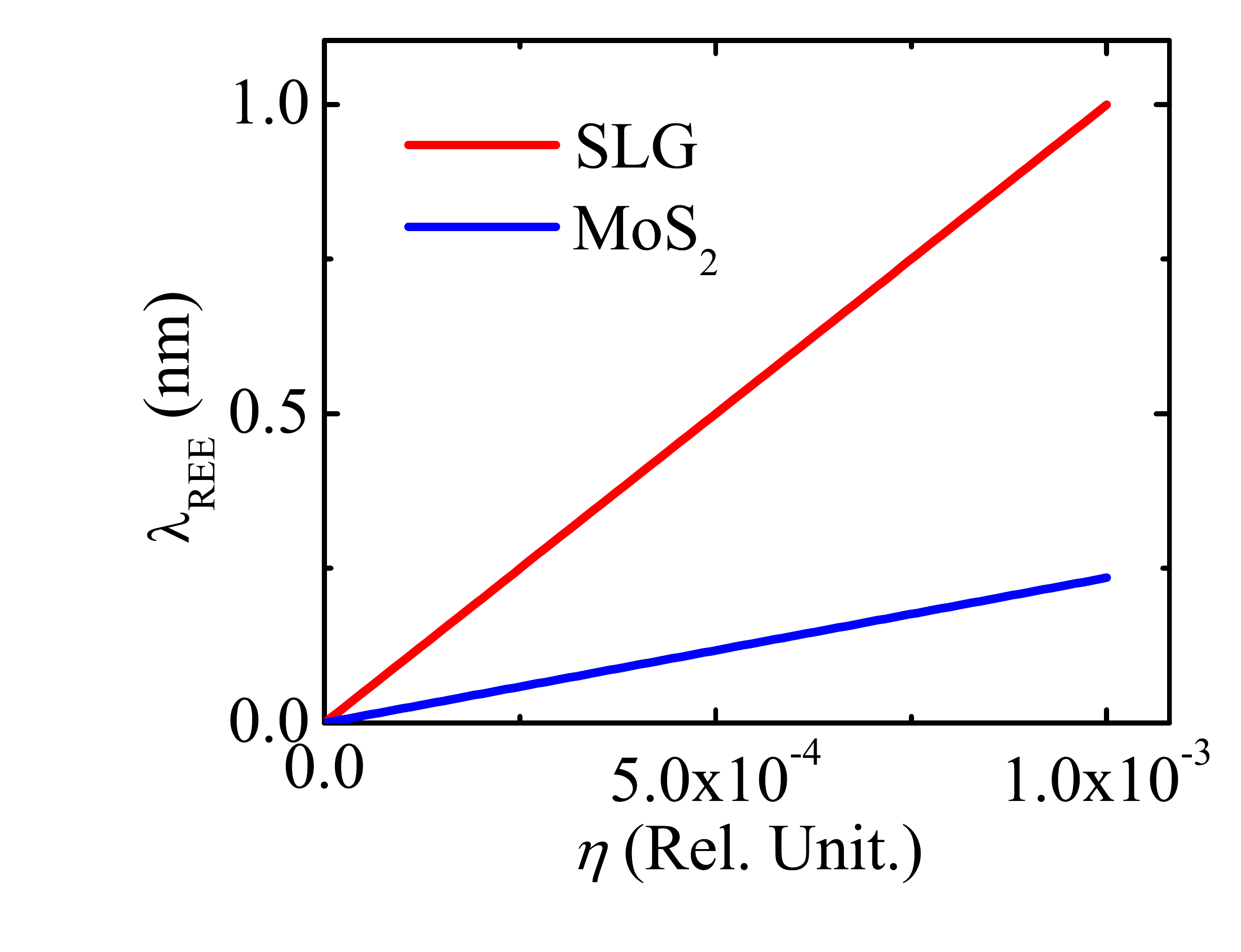}
		\caption{\label{arttype}(online color) Shown REE length $\lambda_{REE}$ (nm) as a function of the REMR $\eta$. The points (red-SLG and green-MoS$_2$) were obtained with the spin diffusion lengths of  $\lambda_{SLG} = 1.0 \times 10^{-6}$ m [25, 34] and $\lambda_{MoS_2} = 235 \times 10^{-9}$ m [16, 34, 35], respectively}
		\label{parede}
	\end{center}
\end{figure}

In accords with Fig. \ref{gato} the REE is an effect of the order of $\lambda_{REE}^{2}$ that becomes relevant only when $l_{2D}$ is sufficiently small. Now is important discuss the limit in which the spin current transverse due the spin accumulation to $\hat{m}$ is completely absorbed as an spin transfer torque without reflection $\Gamma = g_{r}\gg1/(\lambda_{2D}\lambda_{REE}R_{2D})$, which occurs in 2D interface. The spin current at the interface is then 

\begin{equation}
	\frac{J_{S}^{(FM)}}{J_{SO}^{REE}} =
	\stackrel{\Gamma}{\longrightarrow} = \left[tanh\left(\frac{l_{2D}}{\lambda_{2D}}\right)tanh\left(\frac{l_{2D}}{2\lambda_{2D}}\right)\right]\hat{m} \times (\hat{m} \times \hat{y}),
	\label{27}
\end{equation}
and the maximum magnetoresistance for the FM/2DM structure is

\begin{equation}
	\frac{\Delta R_{2D}^{(1)}}{R_{2D}} = \left(\frac{2\lambda_{REE}}{l_{2D}}\right)^2\left(\frac{\lambda_{2D}}{l_{2D}}\right)tanh\left(\frac{l_{2D}}{\lambda_{2D}}\right)tanh^{2}\left(\frac{l_{2D}}{2\lambda_{2D}}\right),
	\label{28}
\end{equation}
but for small thickness (2D interface) $l_{2D} \ll \lambda_{2D}$, we have 

\begin{equation}
	\lambda_{REE} = \lambda_{2D} \eta,
	\label{29}
\end{equation}
where $\eta = (\Delta R_{2D}^{(1)}/R_{2D})^{1/2}$. In Fig. \ref{parede}, it is shown REE length $\lambda_{REE}$ (nm) as a function of the REMR $\eta$ for SLG and MoS$_2$, with spin diffusion length of $\lambda_{SLG} = 1.0 \times 10^{-6}$ m \cite{Mori,Edelstein} and $\lambda_{MoS_2} = 235 \times 10^{-9}$ m \cite{Aparecido, Edelstein, Djeffal}, respectively.      

\subsection{Experimental applications}

\subsubsection{YIG/SLG}
\subsubitem

The SLG has been considered to be very promising materials for spintronic applications \cite{YWangY,McCreary,Pesin,Alves,Mori}. However, due to the low atomic number of carbon, intrinsic graphene has a weak SOC and thus very small spin Hall effect \cite{Mori}. SLG have 2D electronic states that are expected to exhibit remarkable pseudospin. This gives rise to a proximity effect that results in long-range ferromagnetic ordering in graphene, as observed in YIG/SLG \cite{Leutenantsmeyer, Mori, Djeffal}. In fact, the SLG on the YIG film represent one excellent example for application of the study proposal here. For SLG, it was possible to consider the effective thickness $l_{SLG} = 2 \times 10^{-10}$ m, $\Delta R_{2D}^{(1)}/R_{2D} = 0.5 \times 10^{-8}$ and the spin diffusion length $\lambda_{SLG} = 1.0 \times 10^{-6}$ m as in Ref. \cite{Mori, Djeffal}. Then, using the equation (29) is obtain for graphene REE length $\lambda_{REE} = 0.7 \times 10^{-10}$ m, which is in accord with the value measured with electric spin pumping experiments \cite{Mori}. In Fig. \ref{tatu}, it is shown the REMR $\Delta R_{2D}^{(1)}/R_{2D}$ as a function of graphene REE length $\lambda_{REE}$ (nm). The point in red was measured in ref. \cite{Djeffal}.          

\subsubsection{YIG/MoS$_2$}
\subsubitem

\begin{figure}[h]
	\vspace{0.5mm} \hspace{0.5mm}
	\begin{center}
		\includegraphics[scale=0.31]{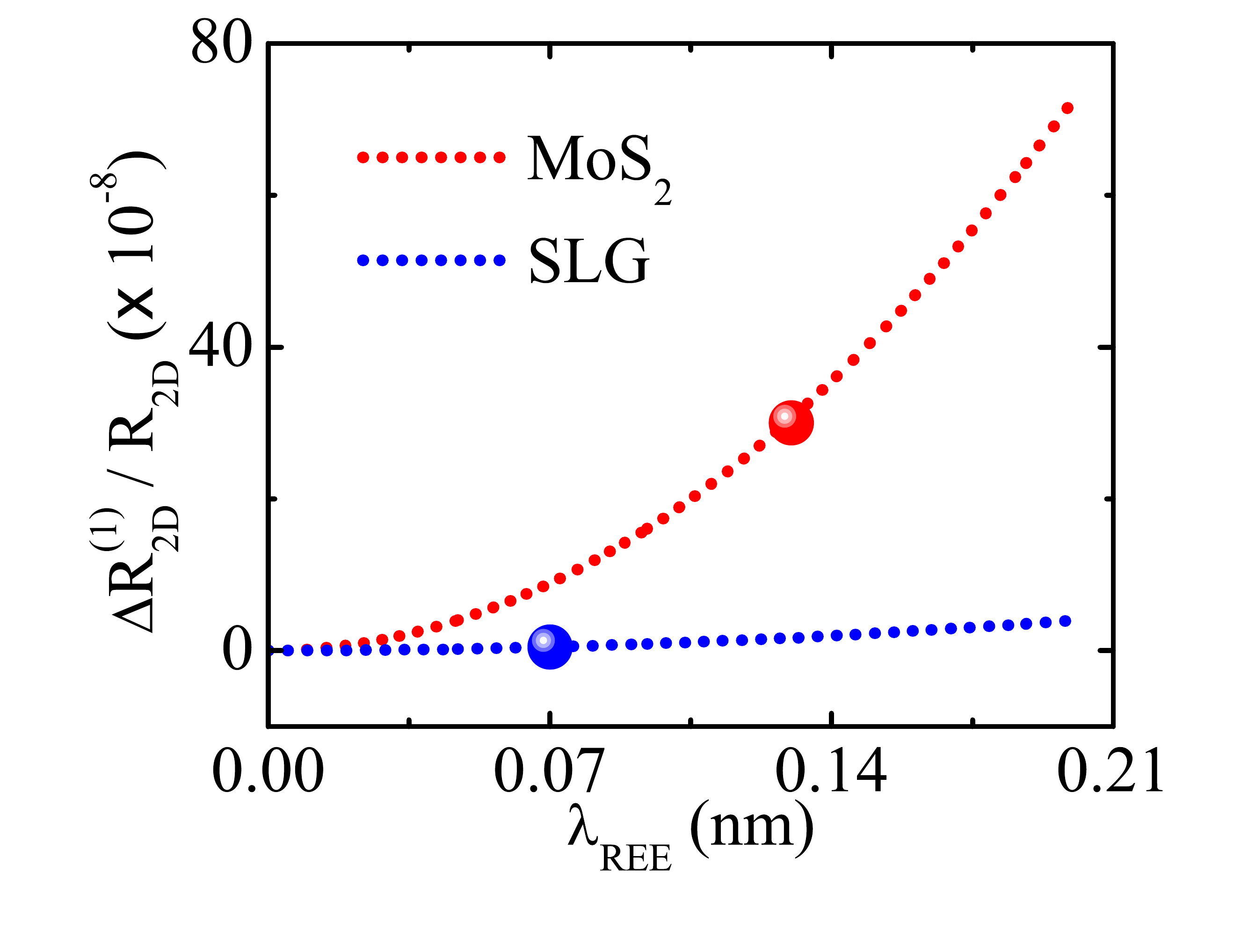}
		\caption{\label{arttype}(online color) REMR $\Delta R_{2D}^{(1)}/R_{2D}$ versus REE length $\lambda_{REE} (10^{-9}$ m) for graphene and MoS$_2$. The points (red-graphene and green-MoS$_2$) were measured in reference \cite{Edelstein}.}
		\label{tatu}
	\end{center}
\end{figure}

Several materials in the family of transition metal dichalcogenides (TMDs) \cite{Gong,Zhong,Mak,Palacios,Amani,Aparecido,Zhang,Shao,Almeida} can also be used for spin-charge current conversion \cite{Aparecido,Edelstein}. Due to their layered structure, the TMD can be easily prepared with one or several atomic layers as to tailor the transport properties. One important TMD material, molybdenum disulfide (MoS$_2$), has attracted widespread attention for a variety of next-generation electrical and optoelectronic device applications because of its unique properties \cite{Gong,Zhong,Mak,Palacios,Amani,Aparecido,Zhang,Shao,Almeida}. For MoS$_2$ it, was used the thickness $t_{MoS_2} = 2.4 \times 10^{-9}$ m, the spin diffusion length  $\lambda_{MoS_2} = 235 \times 10^{-9}$ m \cite{Aparecido,Edelstein,Djeffal} and the REMR, $\Delta R_{2D}^{(1)}/R_{2D} = 30 \times 10^{-8}$. Hence, for YIG/MoS$_2$ we obtain with equation (29) one REE length of $\lambda_{REE} = 0.13
\times 10^{-9}$ m, which is also in good agreement with value measured with electric spin pumping experiments \cite{Aparecido}. In Fig. \ref{tatu}, it is shown the REMR $\Delta R_{2D}^{(1)}/R_{2D}$ as a function of MoS$_2$ REE length $\lambda_{REE}$ (nm). The point in green was measured in ref. \cite{Edelstein}.      

\subsubsection{Exchange field acting on the spin accumulation}

\subsubitem

One ferromagnetic material in atomic contact with 2D material generates a exchange field. The exchange-coupling is caracterized here by $H_{2D}^{Exc} = E_{2D}^{Exc} g_{i} /2e$, which was obtained using the equations (5) and (6). The term $E_{2D}^{Exc}$ is the exchange energy, which for the YIG/2DM interface is $E_{2D}^{Exc} = 1.92 \pm 0.96 \times 10^{-20}$ J (or $E_{2D}^{Exc} = 0.12 \pm 0.06$ eV) \cite{Aparecido, Edelstein, JSMRezende, Ghazaryan, GPWei, Fanchiang}. For the YIG/SLG strtucture \cite{Alves, Mori, Edelstein} with resistance of $R_{2D} = 9 \times 10^3$ $\Omega$, the imaginary part of the spin-mixing interface conductance is of the order of $g_{i} = 4.4 \times 10^8$ m$^{-1} \Omega^{-1}$, thus the exchange field finded was of $H_{2D}^{Exc} = 2.64 \pm 1.32 \times 10^{7}$ A/m (or $\mu_{0}H_{2D}^{Exc} = 33.2 \pm 16.6$ T). Already for the YIG/MoS$_{2}$ structure \cite{Aparecido,Edelstein}, the imaginary part of the spin-mixing interface conductance is of the order of $g_{i} = 2.4 \times 10^7$ m$^{-1} \Omega^{-1}$. In this case, the exchange field obtained was of $H_{2D}^{Exc} = 1.44 \pm 0.72 \times 10^{6}$ A/m (or $\mu_{0}H_{2D}^{Exc} = 1.8 \pm 0.9$ T). The intensity of exchange field acting on the spin accumulation is of the order of exchange field due to the proximity effect obtained by others methods \cite{Ghazaryan, GPWei}.

\section{Conclusion}

In summary, we present a study that describes the Rashba-Edelstein magnetoresistance in 2D materials. The study was applied the measures of REMR makes at room temperature in single layer graphene and in the 2D semiconductor MoS$_2$ in contact with the ferrimagnetic insulator yttrium iron garnet (YIG) measured by the modulated magnetoresistance technique. In the presented discussion, the change in the electrical resistance is reminiscent of the magnetoresistance despite the fact that 3D SOC is not responsible for the magnetoresistance in 2DM. Furthermore, the measured REE lengths for these two materials are in good agreement with the study, this is, which represents a good validation for the present analytical proposal, opening one new method to study the REMR.

\section*{Acknowledgements}

This research was supported by the Brazilian National Council for Scientific and Technological Development (CNPq), Coordination for the Improvement of Higher Education Personnel - Federal Rural University of Pernambuco (CAPES-UFRPE), Financier of Studies and Projects (FINEP) and Foundation for Support to Science and Technology of the State of Pernambuco (FACEPE).

\section*{Author Contributions}

All authors contributed to the study conception and design.

\section*{Data availability statement}

The data will be made available on reasonable request.

\section*{Conflicts of interest}

All the authors declare that there is no conflict of interest.

\bibliographystyle{MiKTeX}

\section*{References}
\begin{thebibliography}{35}
	\bibitem{Soumyanarayanan}{A. Soumyanarayanan, N. Reyren, A. Fert, and C. Panagopoulos, Emergent phenomena induced by spin-orbit coupling at surfaces and interfaces, Nature (London) \textbf{539}, 509 (2016).}
	\bibitem{Hellman}{F. Hellman, A. Hoffmann, Y. Tserkovnyak, G. S. D. Beach, E. E. Fullerton, C. Leighton, A. H. MacDonald, D. C. Ralph, D. A. Arena, H. A. Dürr et al., Interface-induced phenomena in magnetism, Rev. Mod. Phys. \textbf{89}, 025006 (2017).}
	\bibitem{Han}{W. Han, Y. Otani, and S. Maekawa, Quantum materials for spin and charge conversion, NPJ Quantum Mater. \textbf{3}, 27 (2018).}
	\bibitem{Rojas}{J. C. Rojas-Sánchez and A. Fert, Compared efficiencies of conversions between charge and spin current by spin-orbit interactions in two- and three-dimensional systems, Phys. Rev. Appl. \textbf{11}, 054049 (2019).}
	\bibitem{Leutenantsmeyer}{J. C. Leutenantsmeyer, A. A. Kaverzin, M. Wojtaszek and B. J. van Wees. Proximity induced room temperature ferromagnetism in graphene probed with spin currents. 2D Materials, \textbf{4}, 014001 (2017).}
	\bibitem{YWangY}{Y. Wang, Y. Huang, Y. Song, X. Zhang, Y. Ma, J. Liang, and Y. Chen, Room temperature ferromagnetism of graphene, Nano Lett. \textbf{9}, 220 (2008).}
	\bibitem{McCreary}{K. M. McCreary, A. G. Swartz, W. Han, J. Fabian, and R. K. Kawakami, Magnetic Moment Formation in Graphene Detected by Scattering of Pure Spin Currents, Phys. Rev. Lett. \textbf{109}, 186604 (2012).}
	\bibitem{Chang}{C.-Z. Chang, J. Zhang, X. Feng, J. Shen, Z. Zhang, M. Guo, K. Li, Y. Ou, P. Wei, L.-L. Wang et al., Experimental observation of the quantum anomalous Hall effect in a magnetic topological insulator, Science \textbf{340}, 167 (2013).}
	\bibitem{Wei}{P. Wei, F. Katmis, B. A. Assaf, H. Steinberg, P. Jarillo-Herrero, D. Heiman, and J. S. Moodera, Exchange-coupling-induced symmetry breaking in topological insulators, Phys. Rev. Lett. \textbf{110}, 186807 (2013).}
	\bibitem{Wang}{Z. Wang, C. Tang, R. Sachs, Y. Barlas, and J. Shi, Proximity Induced Ferromagnetism in Graphene Revealed by the Anomalous Hall Effect, Phys. Rev. Lett. \textbf{114}, 016603 (2015).}
	\bibitem{Tang}{C. Tang, Z. Zhang, S. Lai, Q. Tan, and W.-B. Gao, Magnetic proximity effect in graphene/CrBr$_3$ van der Waals heterostructures, Adv. Mater. \textbf{32}, 1908498 (2020).}
	\bibitem{Gong}{C. Gong and X. Zhang, Two-dimensional magnetic crystals and emergent heterostructure devices, Science \textbf{363}, eaav4450 (2019).}
	\bibitem{Zhong}{D. Zhong, K. L. Seyler, X. Linpeng, R. Cheng, N. Sivadas, B. Huang, E. Schmidgall, T. Taniguchi, K. Watanabe, M. A. McGuire et al., Van der Waals engineering of ferromagnetic semiconductor heterostructures for spin and valleytronics, Sci. Adv. \textbf{3}, e1603113 (2017).}
	\bibitem{Mak}{K. F. Mak, C. Lee, J. Hone, J. Shan, and T. F. Heinz, Atomically Thin MoS2: A New Direct-Gap Semiconductor, Phys. Rev. Lett. \textbf{105}, 136805 (2010).}
	\bibitem{Palacios}{W. S. Paz and J. J. Palacios, A theoretical study of the electrical contact between metallic and semiconducting phases in monolayer MoS$_2$, 2D Mater. \textbf{4}, 015014 (2016).}
	\bibitem{Amani}{M. Amani et al., Near-unity photoluminescence quantum yield in MoS$_2$, Science \textbf{350}, 1065 (2015).}
	\bibitem{Aparecido}{J. B. S. Mendes, A. Aparecido-Ferreira, J. Holanda, A. Azevedo, and S. M. Rezende, Efficient Spin to Charge Current Conversion in the 2D Semiconductor MoS$_2$ by Spin Pumping from Yttrium Iron Garnet, Appl. Phys. Lett. \textbf{112}, 242407 (2018).}
	\bibitem{Zhang}{W. Zhang, J. Sklenar, B. Hsu, W. Jiang, M. B. Jungfleisch, J. Xiao, F. Y. Fradin, Y. Liu, J. E. Pearson, J. B. Ketterson, Z. Yang, and A. Hoffmann, Perspective: Interface generation of spin-orbit torques featured, APL Mater. \textbf{4}, 032302 (2016).}
	\bibitem{Shao}{Q. Shao, G. Yu, Y.-W. Lan, Y. Shi, M.-Y. Li, C. Zheng, X. Zhu, L.-J. Li, P. K. Amiri, and K. L. Wang, Strong rashba-edelstein effect-induced spin-orbit torques in monolayer transition metal dichalcogenide/ferromagnet bilayers, Nano Lett. \textbf{16}, 7514 (2016).}
	\bibitem{Almeida}{D. B. de Araújo, R. Q. Almeida, A. C. Gadelha, N. P. Rezende, F. C. C. S. Salomão, F. W. N. Silva, L. C. Campos, and E. B. Barros, Controlling the electronic bands of a 2D semiconductor by force microscopy, 2D Mater. \textbf{7}, 045029 (2020).}
	\bibitem{Pesin}{D. Pesin and A. MacDonald, Spintronics and pseudospintronics in graphene and topological insulators, Nat. Mater. \textbf{11}, 409 (2012).}
	\bibitem{Alves}{J. B. S. Mendes, O. Alves Santos, L. M. Meireles, R. G. Lacerda, L. H. Vilela-Leão, F. L. A. Machado, R. L. Rodríguez-Suárez, A. Azevedo, and S. M. Rezende, Spin current to charge-current conversion and magnetoresistance in a hybrid structure of graphene and yttrium iron garnet, Phys. Rev. Lett. \textbf{115}, 226601 (2015).}
	\bibitem{Vidyasagar}{R. Vidyasagar, O. Alves Santos, J. Holanda, R. O. Cunha, F. L. A. Machado, P. R. T. Ribeiro, A. R. Rodrigues, J. B. S. Mendes, A. Azevedo, and S. M. Rezende. Giant Zeeman shifts in the optical transitions of yttrium iron garnet thin films. Appl. Phys. Lett. \textbf{109}, 122402 (2016).}
	\bibitem{Sanchez}{J. C. Rojas-Sanchez, L. Vila, G. Desfonds, S. Gambarelli, J. P. Attane, J. M. De Teresa, C. Magen, and A. Fert, Spin-to-charge conversion using Rashba coupling at the interface between nonmagnetic materials, Nat. Commun. \textbf{4}, 2944 (2013).}
	\bibitem{Loreto}{J. B. S. Mendes, O. Alves Santos, J. Holanda, R. P. Loreto, C. I. L. de Araujo, C.-Z. Chang, J. S. Moodera, A. Azevedo, and S. M. Rezende, Dirac-surface-state-dominated spin to charge current conversion in the topological insulator (Bi$_{0.22}$Sb$_{0.78}$)$_2$Te$_3$ films at room temperature, Phys. Rev. \textbf{B 96}, 180415(R) (2017).}
	\bibitem{Mori}{J. B. S. Mendes, O. Alves Santos, T. Chagas, R. MagalhãesPaniago, T. J. A. Mori, J. Holanda, L. M. Meireles, R. G. Lacerda, A. Azevedo, and S. M. Rezende, Direct detection of induced magnetic moment and efficient spin-to-charge conversion in graphene/ferromagnetic structures, Phys. Rev. \textbf{B 99}, 214446 (2019).}
	\bibitem{Holanda}{J. Holanda, H. Saglam, V. Karakas, Z. Zang, Y. Li, R. Divan, Y. Liu, O. Ozatay, V. Novosad, J. E. Pearson, and A. Hoffmann, Magnetic damping modulation in IrMn$_3$/Ni$_{80}$Fe$_{20}$ via the magnetic spin Hall effect, Phys. Rev. Lett. \textbf{124}, 087204 (2020).}
	\bibitem{Chen}{S. Y. Huang, X. Fan, D. Qu, Y. P. Chen, W. G. Wang, J. Wu, T. Y. Chen, J. Q. Xiao, and C. L. Chien, Transport magnetic proximity effects in platinum. Phys. Rev. Lett. \textbf{109}, 107204 (2012).}
	\bibitem{Mckenzie}{R. H. Mckenzie, J. S. Qualls, S. Y. Han, and J. S. Brooks, Violation of Kohler’s rule by the magnetoresistance of a quasi two-dimensional metal, Phys. Rev. \textbf{B 57}, 11854 (1998).}
	\bibitem{JHolanda}{J. Holanda, Analyzing the magnetic interactions in nanostructures that are candidates for applications in spintronics, J. Phys. D: Appl. Phys. \textbf{54}, 245004 (2021).}
	\bibitem{Nakayama}{H. Nakayama et al., Spin Hall magnetoresistance induced by a nonequilibrium proximity effect, Phys. Rev. Lett. \textbf{110}, 206601 (2013).}
	\bibitem{Althammer}{M. Althammer, S. Meyer, H. Nakayama, M. Schreier, S. Altmannshofer, M. Weiler, H. Huebl, S. Geprägs, M. Opel, R. Gross et al., Quantitative study of the spin Hall magnetoresistance in ferromagnetic insulator/normal metal hybrids, Phys. Rev. \textbf{B 87}, 224401 (2013).}
	\bibitem{Takahashi}{Y.-T. Chen, S. Takahashi, H. Nakayama, M. Althammer, S. T. B. Goennenwein, E. Saitoh, and G. E. W. Bauer, Theory of spin Hall magnetoresistance, Phys. Rev. \textbf{B 87}, 144411 (2013).}
	\bibitem{Conductiond}{V. M. Edelstein. Spin polarization of conduction electrons induced by electric current in two-dimensional asymmetric electron systems. Solid State Communications, \textbf{233}, 73 (1990).}
	\bibitem{Nomura}{H. Nakayama, Y. Kanno, H. An, T. Tashiro, S. Haku, A. Nomura, and K. Ando, Rashba-Edelstein Magnetoresistance in Metallic Heterostructures, Phys. Rev. Lett. \textbf{117}, 116602 (2016).}
	\bibitem{Adachi}{Uchida, K., Xiao, J., Adachi, H. et al. Spin Seebeck insulator. Nature Mater \textbf{9}, 894–897 (2010).}
	\bibitem{Edelstein}{J. B. S. Mendes, S. M. Rezende, and J. Holanda. Rashba-Edelstein magnetoresistance in two-dimensional materials at room temperature. Phys. Rev. \textbf{B 104}, 014408 (2021).}
	\bibitem{Djeffal}{S. Liang, H. Yang, P. Renucci, B. Tao, P. Laczkowski, S. McMurtry, G. Wang, X. Marie, J.-M. George, S. Petit-Watelot, A. Djeffal, S. Mangin, H. Jaffrès, and Y. Lu, Electrical spin injection and detection in molybdenum disulfide multilayer channel, Nat. Commun. \textbf{8}, 14947 (2017).}
	\bibitem{Jia}{X. Jia, K. Liu, K. Xia and G. E. W. Bauer. Spin transfer torque on magnetic insulators. EPL (Europhysics Letters), 96, 17005 (2011).}	
	\bibitem{JSMRezende}{S. M. Rezende, R. L. Rodríguez-Suarez, and A. Azevedo. Magnetic relaxation due to spin pumping in thick ferromagnetic films in contact with normal metals. Phys. Rev. \textbf{B 88}, 014404 (2013).}
	\bibitem{Ghazaryan}{D. Ghazaryan, M. T. Greenaway, Z. Wang et al. Magnon-assisted tunnelling in van der Waals heterostructures based on CrBr3. Nat Electron 1, 344–349 (2018).}
	\bibitem{GPWei}{P. Wei, S. Lee, F. Lemaitre et al. Strong interfacial exchange field in the graphene/EuS heterostructure. Nature Mater 15, 711–716 (2016).}
	\bibitem{Fanchiang}{Y. T. Fanchiang, K. H. M. Chen, C. C. Tseng et al. Strongly exchange-coupled and surface-state-modulated magnetization dynamics in Bi2Se3/yttrium iron garnet heterostructures. Nat Commun \textbf{9}, 223 (2018).}
	
\end{thebibliography}

\end{document}